\documentstyle[12pt]{article}
\begin{document}
\begin{center}
{\Large\bf On a Generalized $D$-Dimensional Oscillator:}

\vspace{0.5cm}
{\Large\bf  Interbasis Expansions}
\end{center}

\vspace{0.5cm}
\begin{center}
{\bf Ye.M.~Hakobyan
\footnote{yera@thsun1.jinr.ru},
{G.S.~Pogosyan}
\footnote{pogosyan@thsun1.jinr.dubna.su},
and A.N.~Sissakian}
\end{center}

\begin{center}
{Bogoliubov Laboratory of Theoretical Physics,}
{Joint Institute for Nuclear Research,}\\
{141980 Dubna, Moscow region, Russia}
\end{center}

\vspace{1cm}
\begin{abstract}

This article deals with  nonrelativistic study of a $D$-dimensional
superintegrable system, which generalizes the ordinary isotropic
oscillator system. The coefficients for the expansion between the
hyperspherical and Cartesian bases (transition matrix), and vice-versa,
are found in terms of the SU(2) Clebsch--Gordan coefficients
analytically continued to real values of their arguments. The diagram
method, which allow one to construct a transition matrix for arbitrary
dimension, is developed.

\end{abstract}

\baselineskip 0.8 true cm

\section{Introduction}

The present article represents a generalization of the work
done by one of the authors (G.P.) fifteen year ago together
with V.M.Ter-Antonyan and Ya.A.Smorodinsky \cite{PST}.
It could also be  considered as a continuation of the work \cite{HKPS}
published in these Proceedings.  Here we study the superintegrable
$D$ - dimensional oscillator system~\cite{EV1}
corresponding to the singular potential
\begin{equation}
\label{I1}
V = \frac{1}{2} \sum_{i=1}^{D} \left(
{\Omega^2} x^2_i + \frac{k_i^2 - \frac{1}{4}}{x_i^2} \right)
\end{equation}
The constants $\Omega> 0$ and $k_i$ ($i = 1, 2,...,D$) are chosen
to be strictly positive. In the case of  $k_i = 1/2$, eq.~(\ref{I1})
yields the well-known oscillator potential. The Schr\"odinger
equation for the $D$- dimensional potential (\ref{I1})
\begin{eqnarray}
\label{SCH1}
\left[
-\frac{1}{2}\Delta + \frac{1}{2} \sum_{i=1}^{D}
\left({\Omega^2} x^2_i + \frac{k_i^2
- \frac{1}{4}}{x_i^2} \right) \right] \Psi = E \Psi
\end{eqnarray}
is separable in the general ellipsoidal coordinates \cite{HW}
and, particularly, in the Cartesian and polyspherical coordinates.
The potential (\ref{I1}) for $D=2$ and $D=3$ belongs to  the potentials,
which were systematically investigated by professor Ya.A.Smorodinsky
with collaborators in \cite{MSVW1,MSVW2} and later
considered from different point of view (quantization in different
systems of coordinates, path integral treatment, invariant and
noninvariant algebra, quadratic algebra, interbasis expansion)
in \cite{HKPS},[7-13].

The plan of this article is as follows. Section 2 is devoted to
the Schr\"odinger equation for the superintegrable $D$ - dimensional
oscillator in the Cartesian coordinates.
In Section 3, by using the "tree" formalism
\cite{VKS} we construct the hyperspherical basis, which is the solution
of the Schr\"odinger equation in the polyspherical coordinates.
Section 4 is the calculation of the interbasis expansion
coefficients between the hyperpsherical and Cartesian bases and
determines the graphical method of constructing the transition matrix.

\section{Solution of the Schr\"odinger equation}

\subsection{Cartesian basis}

The Cartesian wave functions,  vanishing  as $x_i \rightarrow 0$
and $x_i \rightarrow \infty$ ($i=1.2....,D$),
have the following form \cite{EV2,HKPS,GPS,KMP}:
\begin{eqnarray}
\Psi_{\bf n}({\bf x}) = \prod^{D}_{i=1}\psi_{n_i}(x_i, \pm k_i) =
\prod^{D}_{i=1}
 \sqrt\frac{{\Omega}^{1/2} n_i!}
{\Gamma(n_i \pm k_i+1)}
{\rm exp}\left(- \frac{\Omega}{2} x_i^2
\right)\cdot
\nonumber
\\[2mm]
\label{C6}
\cdot\left(\sqrt{\Omega x_i^2}\right)^{\frac{1}{2} \pm k_i }
L_{n_i}^{\pm k_i}(\Omega{x_i}^2)
\end{eqnarray}
where ${\bf x} = (x_1,\dots,x_{D})$, ${\bf n}=(n_1,\dots,n_{D}),
\, n_i\in {{\rm I\!N}}$ and $L_n^\nu(x)$ are associated Lagerre
polynomials ~\cite{BE}. The wave function (\ref{C6}) is normalized
in the domain $[0,\infty)$
\begin{eqnarray}
\label{C3}
\int\limits_{0}^{\infty}
\Psi_{\bf n'}({\bf x})^*
\Psi_{\bf n'}({\bf x}) d{\bf x} =
\frac{1}{2^D} \delta_{{\bf n' n}},
\end{eqnarray}
and the energy spectrum is
\begin{eqnarray}
\label{C7}
E = \Omega \sum_{i=1}^D ({2n_i \pm k_i +1})
= \Omega (2N+D \pm \sum_{i=1}^D k_i),
\end{eqnarray}
where $N= n_1+n_2+ \dots +n_D$ is the principal quantum number.
Note that the positive sign at $k_i$ has to be taken when
$k_i>\frac12$, and both signs, positive and negative, must be
taken into account if $0<k_i\leq\frac12$.

\subsection{Hyperspherical basis}

Consider the system of coordinates
\begin{eqnarray}
\label{H1}
x_i = r {\tilde{x}_i}, \,\,\,\,\,\,
r = \sqrt{x_1^2+ x_2^2+ \dots + x_D^2},
\end{eqnarray}
where the Cartesian coordinates ${\tilde{x}_i}$, $(i=1,2...,D)$
denote the point on the $(D-1)$-dimensional unit sphere $S_{D-1}$:
$\sum\limits_{i=1}^D{\tilde{x}_i}^2 = 1$. Looking for the wave function
$\Psi({\bf x})$ in the form
\begin{eqnarray}
\label{H2}
\Psi({\bf x}) = R(r) Y({\tilde{x}_1},{\tilde{x}_2},
..., {\tilde{x}_D})
\end{eqnarray}
after partial separation of variables in the Schr\"odinger equation
(\ref{SCH1}) we come to the differential equation for the
radial function $R(r)$
\begin{eqnarray}
\label{H3}
\frac{d^2 R}{d r^2}&+& \frac{D-1}{r}\frac{dR}{d r} +
\left\{2 E - \frac{l(l+D-2)}{r^2} - \Omega^2 r^2\right\}
R = 0,
\end{eqnarray}
and to the equation describing the quantum motion on the $S_{D-1}$
sphere with the Rosochatius            potential \cite{MAC}
\begin{eqnarray}
\label{H4}
\left[- \Delta  + \sum_{i=1}^D\frac{k_i^2-\frac14}
{{\tilde{x}_i}^2}\right]\, Y = l(l+D-2)\, Y,
\end{eqnarray}
where $l$ is the hyperspherical separation constant.
Here $\Delta$ is the Laplacian on the sphere and has the form
$$
\Delta = \sum_{1\leq i\leq k}^D L_{ik}^2,
\,\,\,\,\,\,\,
L_{ik} =  {\tilde{x}_i} \partial_{\tilde{x}_k} -
{\tilde{x}_k} \partial_{\tilde{x}_i}
$$
To solve equation (\ref{H4}) via separation of variables
in the polyspherical system of coordinates, we follow the
graphical method, which was developed by Vilenkin, Kuznetsov and
Smorodinsky in paper \cite{VKS}. According to this method it
is useful to associate a polyspherical system of coordinates with
definite graphs, called  ``tree''. In the $D$- dimensional Euclidean
space with the coordinates ${\tilde{x}_1}, {\tilde{x}_2}, \dots,
{\tilde{x}_D}$ any tree has $D$ free points and $D-1$ nodes.
To each node we ascribe a spherical angle $\theta_i$
($i=1,2, \dots, D-1$) and for each line,  opened (free) or closed,
which goes to the right or left side, we  write a function
$\sin\theta_i$ or $\cos\theta_i$ respectively. In this case, the
coordinate ${\tilde{x}_i}$ may be represented as a product of all
the lines coming toward itself. For example, to the tree on {\it Fig.1}
there correspond the following polyspherical coordinates:
\begin{eqnarray*}
\begin{array}{ll}
{\tilde x_1} = \cos\theta_1\cos\theta_2, &\qquad
{\tilde x_2} = \cos\theta_1\sin\theta_2\cos\theta_3,\\
{\tilde x_3} = \cos\theta_1\sin\theta_2\sin\theta_3, &\qquad
{\tilde x_4} = \sin\theta_1\cos\theta_4\cos\theta_5, \\
{\tilde x_5} = \sin\theta_1\cos\theta_4\sin\theta_5 &\qquad
{\tilde x_6} = \sin\theta_1\sin\theta_4,
\end{array}
\end{eqnarray*}

\vspace{1.3cm}
\hspace{7.5cm}
\unitlength=1.00mm
\special{em:linewidth 0.4pt}
\linethickness{0.4pt}
\begin{picture}(35.00,103.00)
\put(-45.00,100.00){\line(4,-5){40.00}}
\put(-5.00,50.00){\line(4,5){40.00}}
\put(-10.00,100.00){\line(-3,-5){15.00}}
\put(0.00,100.00){\line(3,-5){15.00}}
\put(21.00,100.00){\line(-4,-5){12.00}}
\put(-45.00,103.00){\makebox(0,0)[cc]{$\tilde{x}_1$}}
\put(-10.00,103.00){\makebox(0,0)[cc]{$\tilde{x}_3$}}
\put(0.00,103.00){\makebox(0,0)[cc]{$\tilde{x}_4$}}
\put(21.00,103.00){\makebox(0,0)[cc]{$\tilde{x}_5$}}
\put(35.00,103.00){\makebox(0,0)[cc]{$\tilde{x}_6$}}
\put(9.50,88.50){\oval(3.00,3.00)[t]}
\put(15.50,78.50){\oval(3.00,3.00)[t]}
\put(-4.50,53.50){\oval(3.00,3.00)[t]}
\put(-25.50,78.50){\oval(3.00,3.00)[t]}
\put(-26.00,83.00){\makebox(0,0)[cc]{$\theta_2$}}
\put(15.00,83.00){\makebox(0,0)[cc]{$\theta_4$}}
\put(10.00,93.00){\makebox(0,0)[cc]{$\theta_5$}}
\put(-5.00,58.00){\makebox(0,0)[cc]{$\theta_1$}}
\put(30.00,50.00){\makebox(0,0)[cc]{$Fig.1$}}
\put(-5.00,50.00){\circle*{2.00}}
\put(15.00,75.00){\circle*{2.00}}
\put(-25.00,75.00){\circle*{2.00}}
\put(9.00,85.00){\circle*{2.00}}
\put(-18.00,86.00){\line(-3,4){10.67}}
\put(-18.50,89.50){\oval(3.00,3.00)[t]}
\put(-19.00,94.00){\makebox(0,0)[cc]{$\theta_3$}}
\put(-18.00,86.00){\circle*{2.00}}
\put(-29.00,103.00){\makebox(0,0)[cc]{$\tilde{x}_2$}}
\end{picture}

\vspace{-3.8cm}

To construct the separated solution
\begin{eqnarray}
Y (\tilde{x}_1, \tilde{x}_2, \dots, \tilde{x}_D)  =
\prod_{k=1}^{D-1} f_{k}(\theta_k)
\end{eqnarray}
for equation (\ref{H4}), we follow the paper \cite{VKS} and
introduce four types of vertices or elementary "cells" on a tree,
as illustrated in the first line of {\it Table 1}.

Consider the general cell d) with two closed endpoints on the
first line of {\it Table 1}. Let  $l_s, l_r$ and $l$ be
 separation constants corresponding to the nodes on the cell d)
and the parameters $v_s$ and $v_r$ represent numbers of nodes
above  the origin of the cell to the left and right sides,
respectively. Then, the separated equation corresponding to the angle
$\theta = \theta_d$ is
\begin{eqnarray}
\label{GEN}
\Biggl[\frac{1}{\cos^{v_s}\theta \sin^{v_r}\theta}
\frac{d}{d\theta}{\cos^{v_s}\theta \sin^{v_r}\theta}
\frac{d}{d\theta}
+ l(l+v_r+v_s)
\nonumber\\[2mm]
- \frac{l_s(l_s+v_s-1)}{\cos^2\theta}
- \frac{l_r(l_r+v_r-1)}{\sin^2\theta}\Biggr]
f(\theta) = 0.
\end{eqnarray}
Equation (\ref{GEN}) is of a Peschl-Teller type, and the corresponding
solution \cite{FLUG}, orthonormalized in the region $\theta\in
[0, \frac{\pi}{2}]$, has the following form:
\begin{eqnarray}
\label{H7}
f(\theta)\equiv f_{l_s, v_s; l_r, v_r}^{l}(\theta)=
N_q^{(\alpha_s,\alpha_r)} (\cos\theta)^{l_s}
(\sin\theta)^{l_r} P_q^{(\alpha_r,\alpha_s)} (\cos 2\theta),
\end{eqnarray}
where $q=0,1,2,...$ is a spherical quantum number and
\begin{equation}
\label{H8}
\alpha_r=l_r+\frac{v_r-1}{2}, \,\,\,\,
\alpha_s=l_s+\frac{v_s-1}{2}, \,\,\,\,
2q = l-l_s-l_r.
\end{equation}
The normalization constant is
\begin{equation}
\label{H9}
N^{(\alpha,\beta)}_{q}=
\sqrt{\frac{2(2q+\alpha+\beta+1)\Gamma(q+1)\Gamma(q+\alpha+\beta+1)}
{\Gamma(q+\alpha+1)\Gamma(q+\beta+1)}}.
\end{equation}
The solutions to the separated equations for other cells a), b) and c)
on  {\it Table 1} may be found from eq. (12) by adjusting
to every free endpoint ${\tilde x_i}$: $v_i=0$ and the ``momentum''
$l_i = \frac{1}{2} \pm k_i$. These functions are written in the fourth
line of  {\it Table 1} and have the following form:
\begin{eqnarray}
f_{\pm k_i;\pm k_j}^l(\theta)&=& \frac{1}{2}
f_{\frac{1}{2}\pm k_i,0; \frac{1}{2}\pm k_j,0}^l(\theta)=
\frac{N_q^{(\pm k_i,\pm k_j)}}{2}(\cos\theta)^{\frac{1}{2}\pm k_i}
(\sin\theta)^{\frac{1}{2}\pm k_j} P_q^{(\pm k_j,\pm k_i)} (\cos 2\theta)
\label{H10}
\\[2mm]
f_{\pm k_i; l_r}^l(\theta)&=&\frac{1}{\sqrt{2}}
f_{\frac{1}{2}\pm k_i,0; l_r,v_r}^l(\theta)=
\frac{N_q^{(\pm k_i,\alpha_r)} }{\sqrt{2}}(\cos\theta)^{\frac{1}{2}\pm k_i}
(\sin\theta)^{l_r} P_q^{(\alpha_r,\pm k_i)} (\cos 2\theta)
\label{H11}
\\[2mm]
f_{l_s;\pm k_j}^l(\theta)&=& \frac{1}{\sqrt{2}}
f_{l_s,v_s; \frac{1}{2}\pm k_j,0}^l(\theta)=
\frac{N_q^{(\alpha_s,\pm k_j)} }{\sqrt{2}}(\cos\theta)^{l_s}
(\sin\theta)^{\frac{1}{2}\pm k_j} P_q^{(\pm k_j,\alpha_s)} (\cos 2\theta)
\label{H12}
\end{eqnarray}
Let us go to the radial equation (\ref{H3}), which has the  orthonormalized
solution
\begin{equation}
\label{H5}
R(r)\equiv R_{n_r l}(r) =
\sqrt{\frac{2 \Omega^{l+ D/2}n_r!}{\Gamma(n_r+l+\frac{D}{2})}}
\, {\rm exp}\left(-\frac{\Omega}{2} r^2\right)
\, r^{l} L_{n_r}^{l+\frac{D-2}{2}}(\Omega r^2),
\end{equation}
where $n_r \in {{\rm I\!N}}$ is a radial quantum number and hypermomentum
$l= \sum_{i=1}^{D-1}(q_i \pm k_i + \frac12) + (\frac12 \pm k_{D})$.
The energy spectrum is given by  eq. (5), where the principal quantum
number now is $N = n_r + q_1+q_2+\dots+q_{D-1}$.
The total hyperspherical wave function (7) is given by formulae
(\ref{H5}) and (10)
\begin{equation}
\Psi_{n_r, \,{\bf l}} \, (r, {\vec\theta})  =
R_{n_r l}(r) \, Y_{{\bf l}} \,\, ({\vec\theta}),
\end{equation}
where ${\vec\theta} = (\theta_1,\dots,\theta_{D-1})$, \,
${\bf l} = (l_1,l_2, ...l_{D-1})$, and the connection between
the spherical quantum number $q$ and the separation constants $l_i$
are represented by the fourth line in {\it Table 1}.

\section{Connecting Cartesian and hyperspherical bases}

For the fixed value of energy we can write
the expansion of theCartesian basis $\Psi_{\bf n}({\bf x})$
in terms of the hyperspherical basis $\Psi_{n_r,
{\bf l}}(r, {\vec\theta})$ in the  form
\begin{equation}
\label{CSS1}
{\Psi}_{\bf{n}}({\bf x}) =
\sum_{\bf q} W_{\bf n}^{N, \, {\bf q}}
\, (\pm k_1, \dots, \pm k_N)
\Psi_{n_r, {\bf l}}\, (r, {\vec\theta}).
\end{equation}
Here, the sum is taken over $(D-1)$ quantum numbers
${\bf q}=(q_1, \dots, q_{D-1})$ and determined by the condition
$N = n_1+\dots+n_D=n_r+q_1+\dots+q_{D-1}$.
By multiplying both sides of the expansion (\ref{CSS1}) by the
factor $r^{-D}$ and using the asymptotic formula for the associated
Laguerre polynomials $L_{n}^{\alpha}(x)$ for large $x$
\begin{equation}
L_n (x) \sim \frac{(-x)^n}{n!},
\end{equation}
eq.~(\ref{CSS1}) yields an equation dependent only on variables
${\vec\theta}$. Then, by using the orthogonality property of
$Y_{\bf l}\,({\vec\theta})$ in the region $\theta_i \in [0,
\frac{\pi}{2}]$, we obtain the integral representation for the
transition matrix (\ref{CSS1})
\begin{eqnarray}
\label{CSS2}
W_{\bf n}^{N, {\bf q}} (\pm k_1, \dots, \pm k_N)
= M \times
\int d\Omega({\vec\theta}) Y_{\bf l}\, ({\vec\theta})
\prod_{i=1}^{N}({\tilde x_i})^{\tilde n_i}
\end{eqnarray}
where
\begin{eqnarray}
\label{CSS3}
M = \frac{(-1)^{N-n_r}}{\sqrt{2}}
\sqrt{\frac{2 ^{2D}n_r!\Gamma(n_r+l+\frac{D}{2})}
{\prod^{D}_{i=1}\left[n_i!\Gamma(n_i\pm k_i+1)\right]}}
\end{eqnarray}
and ${\tilde n_i} = (2n_i \pm k_i + \frac{1}{2})$.

To calculate the transition matrix (\ref{CSS2}), we must know exactly
the form of the tree, corresponding to the polyspherical coordinates,
and the contribution from
each of the cells (see {\it Table 1}) to the functions
$d\Omega({\vec\theta})$, $Y_{\bf l}\, ({\vec\theta})$ and
$\prod_{i=1}^{D}({\tilde x_i})^{\tilde n_i}$.
Therefore, the matrix (\ref{CSS2}) includes only four types of
integrals
\begin{eqnarray}
\label{CSS4}
F_{{\tilde n_i}, {\tilde n_j}}^{l}(\pm k_i,\pm k_j)&=&
\int^{\frac{\pi}{2}}_{0}(\cos\theta)^{\tilde n_i}
(\sin\theta)^{\tilde n_j}
f_{\pm k_i;\pm k_j}^l(\theta) d \theta,
\\[2mm]
\label{CSS5}
F_{{\tilde n_i}; N_r}^{l}(\pm k_i; l_r, v_r)&=&
\int^{\frac{\pi}{2}}_{0}(\cos\theta)^{\tilde n_i}
(\sin\theta)^{N_r}
f_{\pm k_i; l_r}^{l}(\theta) d \theta,
\\[2mm]
\label{CSS6}
F_{N_s, {\tilde n_j}}^{l}(L_s,v_s;\pm k_j)&=&
\int^{\frac{\pi}{2}}_{0}
(\cos\theta)^{N_s}(\sin\theta)^{\tilde n_i}
f_{l_s; \pm k_j}^{l}(\theta) d \theta,
\\[2mm]
\label{CSS7}
F_{N_s; N_r}^{l}(l_s, v_s; l_r, v_r)&=&
\int^{\frac{\pi}{2}}_0
(\cos\theta)^{N_s}(\sin\theta)^{N_r}
f_{l_s; l_r}^{l}(\theta) d \theta,
\end{eqnarray}
where  $N_s$ and $N_r$ are the sums of all ${\tilde n_i}$
above the cell
on the left and right sides, respectively.

Let us now  calculate the general integral (\ref{CSS7}). Using
the Rodrigues formula for the Jacobi polynomials~\cite{BE},
we obtain
$$
F_{N_s; N_r}^{l}(l_s,v_s; l_r, v_r)
=
\frac{(-1)^{\frac{l-l_s-l_r}{2}}}
{2^{\frac{N_s+N_r+l_s+l_r+v_s+v_r+2}{2}}}
\sqrt{\frac{(l+\frac{v_s+v_r}{2})
\Gamma(\frac{l+l_s+l_r}{2} + \frac{v_s+v_r}{2})}
{\Gamma(\frac{l-l_s-l_r}{2}+1)\Gamma(\frac{l+l_s-l_r}{2}
+ \frac{v_r+1}{2})
\Gamma(\frac{l-l_s+l_r}{2} + \frac{v_s+1}{2})}}
$$
\begin{eqnarray}
\label{CSS8}
\int^{1}_{-1} (1+x)^{N_s-l_s}
(1-x)^{N_r-l_r}
\frac{d^{\frac{l-l_s-l_r}{2}}}
{d x^{\frac{l-l_s-l_r}{2}}}
\left[(1+x)^{\frac{l+l_s-l_r}{2} + \frac{v_s-1}{2}}
(1-x)^{\frac{l-l_s+l_r}{2} + \frac{v_r-1}{2}}
\right] d x
\end{eqnarray}
Comparing (\ref{CSS8}) with the integral representation of the Clebsh-Gordan
coefficients $C_{a, \alpha; b, \beta}^{c, \gamma}$ for the group
SU(2) ~\cite{VMK}, we obtain
\begin{eqnarray}
\label{CSS9}
F_{N_s; N_r}^{l}(l_s,v_s; l_r, v_r)
=
\frac{(-1)^{q+\frac{l-l_s-N_r}{2}}}{\sqrt{2}}
\,
K_{N_s, N_r}^{l_s, v_s; l_r, v_r}
\,
C_{a, \alpha; b, \beta}^{c, \gamma}
\end{eqnarray}
with
\begin{eqnarray}
\label{CSS10}
K_{N_s, N_r}^{l_s, v_s; l_r, v_r} =
\sqrt{\frac{\Gamma(\frac{N_s-l_s}{2}+1)
\Gamma(\frac{N_s+l_s}{2}+\frac{v_s+1}{2})
\Gamma(\frac{N_r-l_r}{2}+1)
\Gamma(\frac{N_r+l_r}{2}+\frac{v_r+1}{2})}
{\Gamma(\frac{N_r+N_s+l}{2}+\frac{v_r+v_s}{2}+1)
\Gamma(\frac{N_s+N_r-l}{2}+1)}}
\end{eqnarray}
and
\begin{eqnarray*}
4 a&=& {l_s - l_r + N_s + N_r +v_s -1}, \quad
4 b = {l_r - l_s + N_s + N_r +v_r -1},
\\[2mm]
4 \alpha&=& {l_r + l_s + N_s - N_r +v_s -1}, \quad
4 \beta = {l_r + l_s + N_r - N_s +v_r -1},
\\[2mm]
2c &=&l + \frac{v_s-1}{2} + \frac{v_r-1}{2}, \quad
\,\,
2\gamma = l_s+l_r + \frac{v_s-1}{2} + \frac{v_r-1}{2}.
\end{eqnarray*}
By realizing that the integrals (\ref{CSS4})-(\ref{CSS6}) may
be expressed through the integral (\ref{CSS7}), we obtain
\begin{eqnarray}
\label{CSS11}
F_{{\tilde n_i}, {\tilde n_j}}^{l}(k_i, k_j)
&=&
\frac{1}{2}F_{{\tilde n_i}; {\tilde n_j}}^{l}
(\frac{1}{2}\pm k_i,0; \frac{1}{2}\pm k_j, 0),
\\
\label{CSS12}
F_{N_s, {\tilde n_j}}^{l}(l_s, v_s, k_j)
&=&
\frac{1}{\sqrt{2}}F_{N_s; {\tilde n_j}}^{l}
(l_s,v_s; \frac{1}{2}\pm k_j, 0),
\\
\label{CSS13}
F_{{\tilde n_i}, N_r}^{l}(k_i, l_r,v_r)
&=&
\frac{1}{\sqrt{2}}F_{{\tilde n_i}; N_r}^{l}
(\frac{1}{2}\pm k_i,0; l_r, v_r),
\end{eqnarray}
Thus, the contributions from the four types of cells in the matrix
(\ref{CSS2}) are calculated and  given by formulae (\ref{CSS9})
and (\ref{CSS11}) - (\ref{CSS13}).

Let us now construct the transition matrix (\ref{CSS2}).
At first, we need to prove that the matrix (\ref{CSS2})
can be expressed in terms of the product of the
Clebsch-Gordan coefficients  only.
Indeed, let ``$s$'', ``$r$'' and ``$d$''
be the quantum numbers corresponding to the cells  ``$s$'' and ``$r$''
corresponding to the left and right top in this cell and ``$d$''  being  the
origin of the cell. Therefore, using the relation
$v_d = v_s+v_r+1$, we can write
\begin{eqnarray}
\label{CSS14}
K_{N_s, N_r}^{l_s, v_s; l_r, v_r} \equiv K(s,r;d) =
\frac{f(s)f(r)}{f(d)}
\end{eqnarray}
where
\begin{eqnarray}
\label{CSS15}
f(i) = \sqrt{\Gamma\left(\frac{N_i-l_i}{2}+1\right)
\Gamma\left(\frac{N_i+l_i}{2}+\frac{v_i+1}{2}\right)}
\end{eqnarray}
The full contribution of  constants (\ref{CSS10}) to the transition
matrix (\ref{CSS2}) is equal to the product of the constants (\ref{CSS14})
upon all the cells
\begin{eqnarray}
\label{CSS16}
\prod_{i=1}^{D-1} K_i(s,r;d) =
\frac{1}{f(d)} \prod_{i=1}^{D} f(i)
=
\sqrt{\frac{\prod_{j=1}^{D}\Gamma(n_j+1)\Gamma(n_j\pm k_j+1)}
{\Gamma(n_r+l + \frac{D}{2})
\Gamma(n_r+1)}},
\end{eqnarray}
where $N_s+N_r = 2n_r + l$. Then, the contribution of all the
constants (\ref{CSS10}) with the coefficient (23) is eliminated
and for any tree the transition matrix (\ref{CSS2}) has the
following form:
\begin{eqnarray}
\label{CSS18}
W_{\bf n}^{N, \, {\bf q}} (\pm k_1, \dots, \pm k_D )
=\prod_{i=1}^{D-1}
(-1)^{c_i-a_i-\beta_i}
C_{a_i, \alpha_i; b_i, \beta_i}^{c_i, \gamma_i}
\end{eqnarray}
with $a_i, b_i, \alpha_i, \beta_i, c_i, \gamma_i$ given by formula
(\ref{CSS9}), and the multiplication is taken upon all the cells.
The quantum numbers in (\ref{CSS18}) for $k_i\not= 1/2$ are not
integers or half or odd integers and, therefore, the coefficients
in the matrix (\ref{CSS18}) may be considered as an analytic
continuation of the SU(2) Clebsch--Gordan coefficients for the real
values of their arguments.
For $D=2$ and $D=3$
we obtain the result from paper \cite{HKPS}.

Determine now the graphical methods of constructing the matrix
(\ref{CSS18}), which we  call a transition ``tree'' and
which is identical to the corresponding hyperspherical ``tree''.
Let to any free endpoint of the tree there correspond  the ``momentum''
$1/2\pm k_i$ and Cartesian quantum number ${\tilde n_i} = 2n_i \pm
k_i+1$ and  to the nodes there correspond  the separation constants
$l_i$, the numbers $v_i$ and $N_i = \sum {\tilde n_i}$.
Then, after drawing the transition ``tree'' and multiplying
the contributions from all the cells in the tree according
to  {\it Table 2}, we come to the final result in the form
(\ref{CSS18}).

Because of the orthogonality properties for the SU(2) Clebsch--Gordan
coefficients, the inverse expansion could be written as
$$
\Psi_{n_r, {\bf l}}\, (r, {\vec\theta})
=
\sum_{n_1+n_2+...+ n_D=N} W_{\bf n}^{* N, \, {\bf q}}
\, (\pm k_1, \dots, \pm k_N)
{\Psi}_{\bf n}({\bf x})
$$

\section{Conclusion}

One of the main results of this paper is the construction of the
hyperspherical wave function which is the solution of the
Schr\"odinger equation for the motion on the $(D-1)$ - dimensional
sphere for the Rosochatius potential \cite{MAC} and which
generalizes the classical hyperspherical function for $k_i\not= 1/2$
\cite{VKS}.

We have also calculated the transition matrix between the
hyperspherical and Cartesian bases and shown that the Clebsch--Gordan
coefficients entering into this matrix are the analytic continuation
of the SU(2) Clebsch--Gordan coefficients for real values of their
arguments. In addition, we propose the diagram method, the "transition tree",
which allows one to construct a transition matrix for an arbitrary tree.

\section*{Acknowledgements}

We thank Professors M.~Kibler and V.~M.Ter-Antonyan and Dr. L.G.Mardoyan
for helpful discussions.


\newpage
\vspace{0.5cm}
\unitlength=1.00mm
\special{em:linewidth 0.4pt}
\linethickness{0.4pt}
\begin{picture}(159.00,130.00)
\put(0.00,20.00){\line(1,0){159.00}}
\put(0.00,40.00){\line(1,0){159.00}}
\put(0.00,60.00){\line(1,0){159.00}}
\put(0.00,80.00){\line(1,0){159.00}}
\put(0.00,100.00){\line(1,0){159.00}}
\put(0.00,130.00){\line(1,0){159.00}}
\put(20.00,130.00){\line(0,0){0.00}}
\put(20.00,130.00){\line(1,0){100.00}}
\put(32.00,120.00){\line(2,-3){10.00}}
\put(42.00,105.00){\line(2,3){10.00}}
\put(97.00,120.00){\line(2,-3){10.00}}
\put(107.00,105.00){\line(2,3){10.00}}
\put(65.00,120.00){\line(2,-3){10.00}}
\put(75.00,105.00){\line(2,3){10.00}}
\put(130.00,120.00){\line(4,-5){12.00}}
\put(142.00,105.00){\line(3,4){11.33}}
\put(130.00,120.00){\circle*{2.00}}
\put(153.00,120.00){\circle*{2.00}}
\put(97.00,120.00){\circle*{2.00}}
\put(85.00,120.00){\circle*{2.00}}
\put(75.00,50.00){\makebox(0,0)[cc]{$f_{\frac{1}{2}\pm k_i; l_r, v_r}^l(\theta)$}}
\put(108.00,50.00){\makebox(0,0)[cc]{$f_{l_s, v_s; \frac{1}{2}\pm k_j}^l(\theta)$}}
\put(15.00,29.00){\makebox(0,0)[cc]{$l$}}
\put(42.00,29.00){\makebox(0,0)[cc]{$2 q\pm k_i\pm k_j+1$}}
\put(75.00,29.00){\makebox(0,0)[cc]{$2 q + l_r \pm k_i+\frac{1}{2}$}}
\put(108.00,29.00){\makebox(0,0)[cc]{$2 q +l_s \pm k_j+\frac{1}{2}$}}
\put(141.00,29.00){\makebox(0,0)[cc]{$2 q+l_s+l_r$}}
\put(15.00,50.00){\makebox(0,0)[cc]{$Y_{{\bf l}}({\vec\theta})$}}
\put(42.00,50.00){\makebox(0,0)[cc]{$f_{\frac{1}{2}\pm k_i, \frac{1}{2}\pm k_j}^l(\theta)$}}
\put(141.00,50.00){\makebox(0,0)[cc]{$f_{l_s,v_s; l_r,v_r}^{l}(\theta)$}}
\put(15.00,70.00){\makebox(0,0)[cc]{$d\Omega(\theta)$}}
\put(42.00,70.00){\makebox(0,0)[cc]{$d\theta$}}
\put(75.00,70.00){\makebox(0,0)[cc]{$(\sin \theta)^{v_r} d\theta$}}
\put(108.00,70.00){\makebox(0,0)[cc]{$(\cos\theta)^{v_s}d \theta$}}
\put(141.00,70.00){\makebox(0,0)[cc]{$(\cos\theta)^{v_s}(\sin\theta)^{v_r}d\theta$}}
\put(15.00,90.00){\makebox(0,0)[cc]{$\theta$}}
\put(42.00,90.00){\makebox(0,0)[cc]{$0\leq\theta\leq 2\pi$}}
\put(75.00,90.00){\makebox(0,0)[cc]{$0\leq\theta\leq\pi$}}
\put(108.00,90.00){\makebox(0,0)[cc]{$-\frac{\pi}{2}\leq\theta\leq\frac{\pi}{2}$}}
\put(141.00,90.00){\makebox(0,0)[cc]{$0\leq\theta\leq\frac{\pi}{2}$}}
\put(129.00,125.00){\makebox(0,0)[cc]{$l_s, v_s$}}
\put(154.00,125.00){\makebox(0,0)[cc]{$l_r, v_r$}}
\put(34.00,125.00){\makebox(0,0)[cc]{$\frac{1}{2}\pm k_i$}}
\put(50.00,125.00){\makebox(0,0)[cc]{$\frac{1}{2}\pm k_j$}}
\put(67.00,125.00){\makebox(0,0)[cc]{$\frac{1}{2}\pm k_i$}}
\put(83.00,125.00){\makebox(0,0)[cc]{$l_r, v_r$}}
\put(99.00,125.00){\makebox(0,0)[cc]{$l_s, v_s$}}
\put(116.00,125.00){\makebox(0,0)[cc]{$\frac{1}{2}\pm k_j$}}
\put(47.00,105.00){\makebox(0,0)[cc]{$l$}}
\put(81.00,105.00){\makebox(0,0)[cc]{$l$}}
\put(112.00,105.00){\makebox(0,0)[cc]{$l$}}
\put(148.00,105.00){\makebox(0,0)[cc]{$l$}}
\put(75.00,110.50){\oval(4.00,3.00)[t]}
\put(107.00,109.50){\oval(4.00,3.00)[t]}
\put(142.00,109.50){\oval(4.00,3.00)[t]}
\put(42.00,109.50){\oval(4.00,3.00)[t]}
\put(42.00,114.00){\makebox(0,0)[cc]{$\theta_a$}}
\put(75.00,114.00){\makebox(0,0)[cc]{$\theta_b$}}
\put(107.00,114.00){\makebox(0,0)[cc]{$\theta_c$}}
\put(142.00,114.00){\makebox(0,0)[cc]{$\theta_d$}}
\put(26.00,130.00){\line(0,-1){110.00}}
\put(124.00,130.00){\line(0,-1){110.00}}
\put(91.00,130.00){\line(0,-1){110.00}}
\put(128.00,-16.00){\makebox(0,0)[cc]{Table 1.}}
\put(0.00,130.00){\line(0,-1){110.00}}
\put(159.00,130.00){\line(0,-1){110.00}}
\put(75.00,105.00){\circle*{2.00}}
\put(107.00,105.00){\circle*{2.00}}
\put(142.00,105.00){\circle*{2.00}}
\put(42.00,105.00){\circle*{2.00}}
\put(31.00,104.00){\makebox(0,0)[cc]{$a)$}}
\put(64.00,104.00){\makebox(0,0)[cc]{$b)$}}
\put(97.00,104.00){\makebox(0,0)[cc]{$c)$}}
\put(129.00,104.00){\makebox(0,0)[cc]{$d)$}}
\put(0.00,0.00){\line(1,0){159.00}}
\put(124.00,0.00){\line(0,1){20.00}}
\put(91.00,20.00){\line(0,-1){20.00}}
\put(26.00,20.00){\line(0,-1){20.00}}
\put(15.00,9.00){\makebox(0,0)[cc]{$\prod\limits_{i=1}^{D} (\tilde{x}_i)^{\tilde{n}_i}$}}
\put(42.00,9.00){\makebox(0,0)[cc]{$(\cos\theta)^{\tilde{n}_i}(\sin\theta)^{\tilde{n}_j}$}}
\put(75.00,9.00){\makebox(0,0)[cc]{$(\cos\theta)^{\tilde{n}_i}(\sin\theta)^{N_r}$}}
\put(108.00,9.00){\makebox(0,0)[cc]{$(\cos\theta)^{N_s}(\sin\theta)^{\tilde{n}_j}$}}
\put(141.00,9.00){\makebox(0,0)[cc]{$(\cos\theta)^{N_s}(\sin\theta)^{N_r}$}}
\put(58.00,0.00){\line(0,1){130.00}}
\put(0.00,20.00){\line(0,-1){20.00}}
\put(159.00,20.00){\line(0,-1){20.00}}
\end{picture}

\unitlength=1.00mm
\special{em:linewidth 0.4pt}
\linethickness{0.4pt}
\begin{picture}(159.00,120.00)
\put(0.00,120.00){\line(1,0){159.00}}
\put(159.00,120.00){\line(0,0){0.00}}
\put(0.00,90.00){\line(1,0){159.00}}
\put(159.00,65.00){\line(-1,0){159.00}}
\put(0.00,120.00){\line(0,-1){55.00}}
\put(159.00,120.00){\line(0,-1){55.00}}
\put(41.00,120.00){\line(0,-1){55.00}}
\put(80.00,120.00){\line(0,-1){55.00}}
\put(120.00,120.00){\line(0,-1){55.00}}
\put(20.00,97.00){\line(4,5){12.67}}
\put(20.00,97.00){\line(-3,4){12.00}}
\put(60.00,97.00){\line(4,5){12.67}}
\put(60.00,97.00){\line(-3,4){12.00}}
\put(100.00,97.00){\line(4,5){12.67}}
\put(100.00,97.00){\line(-3,4){12.00}}
\put(140.00,97.00){\line(4,5){12.67}}
\put(140.00,97.00){\line(-3,4){12.00}}
\put(73.00,114.00){\circle*{2.00}}
\put(88.00,113.00){\circle*{2.00}}
\put(128.00,113.00){\circle*{2.00}}
\put(153.00,113.00){\circle*{2.00}}
\put(140.00,97.00){\circle*{2.00}}
\put(100.00,97.00){\circle*{2.00}}
\put(60.00,97.00){\circle*{2.00}}
\put(20.00,97.00){\circle*{2.00}}
\put(121.00,116.00){\makebox(0,0)[lc]{$l_s,v_s,N_s$}}
\put(158.00,116.00){\makebox(0,0)[rc]{$l_r,v_r,N_r$}}
\put(142.00,97.00){\makebox(0,0)[lc]{$N_s+N_r,$}}
\put(138.00,92.00){\makebox(0,0)[lc]{$v_s+v_r+1$}}
\put(20.00,82.00){\makebox(0,0)[cc]{$(-1)^{\frac{l-\frac12\mp k_i-\tilde n_j}{2}}C^{c\gamma}_{a\alpha,b\beta}$}}
\put(118.00,116.00){\makebox(0,0)[rc]{$\frac12\pm k_j,\tilde n_j$}}
\put(82.00,116.00){\makebox(0,0)[lc]{$l_s,v_s,N_s$}}
\put(78.00,116.00){\makebox(0,0)[rc]{$l_r,v_r,N_r$}}
\put(43.00,116.00){\makebox(0,0)[lc]{$\frac12\pm k_i,\tilde n_i$}}
\put(1.00,116.00){\makebox(0,0)[lc]{$\frac12\pm k_i,\tilde n_i$}}
\put(40.00,116.00){\makebox(0,0)[rc]{$\frac12\pm k_j,\tilde n_j$}}
\put(20.00,76.00){\makebox(0,0)[cc]{$v_s=v_r=0;$}}
\put(21.00,69.00){\makebox(0,0)[cc]{$l_s=\frac{1}{2}\pm k_i;l_r=\frac{1}{2}\pm k_j$}}
\put(100.00,71.00){\makebox(0,0)[cc]{$v_r=0;l_r=\frac{1}{2}\pm k_j$}}
\put(60.00,71.00){\makebox(0,0)[cc]{$v_s=0;l_s=\frac{1}{2}\pm k_i$}}
\put(24.00,97.00){\makebox(0,0)[lc]{$l,\tilde n_i+\tilde n_j$}}
\put(63.00,97.00){\makebox(0,0)[lc]{$\tilde n_i+N_r,$}}
\put(63.00,93.00){\makebox(0,0)[lc]{$v_r+1$}}
\put(103.00,93.00){\makebox(0,0)[lc]{$v_s+1$}}
\put(103.00,97.00){\makebox(0,0)[lc]{$N_s+\tilde n_j,$}}
\put(126.00,55.00){\makebox(0,0)[cc]{$Table 2$}}
\put(57.00,97.00){\makebox(0,0)[cc]{$l,$}}
\put(97.00,97.00){\makebox(0,0)[cc]{$l,$}}
\put(136.00,97.00){\makebox(0,0)[cc]{$l,$}}
\put(60.00,82.00){\makebox(0,0)[cc]{$(-1)^{\frac{l-\frac12 \mp k_i-N_r}{2}}C^{c\gamma}_{a\alpha,b\beta}$}}
\put(100.00,82.00){\makebox(0,0)[cc]{$(-1)^{\frac{l-l_s-\tilde n_j}{2}}C^{c\gamma}_{a\alpha,b\beta}$}}
\put(140.00,82.00){\makebox(0,0)[cc]{$(-1)^{\frac{l-l_s-N_r}{2}}C^{c\gamma}_{a\alpha,b\beta}$}}
\end{picture}

\end{document}